**Chapter 1**

# VerilogMonkey: Exploring Parallel Scaling for Automated Verilog Code Generation with LLMs

Juxin Niu, Yuxin Du, Dan Niu, Xi Wang, Zhe Jiang, and Nan Guan

**Abstract** We present VerilogMonkey, an empirical study of parallel scaling for the underexplored task of automated Verilog generation. Parallel scaling improves LLM performance by sampling many outputs in parallel. Across multiple benchmarks and mainstream LLMs, we find that scaling to hundreds of samples is cost-effective in both time and money and, even without any additional enhancements such as post-training or agentic methods, surpasses prior results on LLM-based Verilog generation. We further dissect why parallel scaling delivers these gains and show how output randomness in LLMs affects its effectiveness.

Juxin Niu[0009−0004−8223−4245]
Department of Computer Science, City University of Hong Kong, Hong Kong SAR
e-mail: juxin.niu@my.cityu.edu.hk

Yuxin Du[0009−0001−7017−5805]
Department of Computer Science, City University of Hong Kong, Hong Kong SAR
e-mail: yuxindu8-c@my.cityu.edu.hk

Dan Niu[0000−0002−0715−7946]
School of Automation, Southeast University, China
e-mail: 101011786@seu.edu.cn

Xi Wang[0000−0002−1998−6733]
National Center of Technology Innovation for EDA, China
School of Integrated Circuits, Southeast University, China
e-mail: xi.wang@seu.edu.cn

Zhe Jiang[0000−0002−8509−3167]
National Center of Technology Innovation for EDA, China
School of Integrated Circuits, Southeast University, China
e-mail: zhejiang.arch@gmail.com

Nan Guan[0000−0003−3775−911X]
Department of Computer Science, City University of Hong Kong, Hong Kong SAR
e-mail: nanguan@cityu.edu.hk

## 1.1 Introduction

Writing Hardware Description Language (HDL) code such as Verilog is demanding, requires substantial expertise, and is prone to bugs and errors. Therefore, there is growing interest in LLM-assisted hardware code generation [27, 31, 37]. However, the capability of LLMs on automated Verilog code generation is still unsatisfactory, comparing with other tasks such as software code generation.

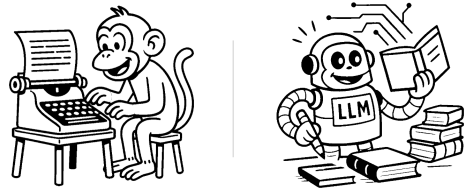

Fig. 1.1: Even randomness can achieve greatness given infinite attempts.[1]

Early progress in improving the capability of LLM was driven by *training-time scaling*, i.e., improving single-shot inference accuracy by training larger models and ingesting ever larger datasets. However, the pace of training-time scaling has gradually slowed due to its resource-intensive nature and the sharply rising costs of human labor [19]. To address this, research attention has shifted to *test-time scaling* [43], which boosts performance by allocating more computational resources during inference. A popular approach in this direction is *parallel scaling* [21, 35], which generates multiple outputs in parallel and aggregates them into a final answer. Parallel scaling has significantly enhanced LLM's performance in tasks such as math [6, 15] and software coding [12, 23]. For example, AlphaCode [24] extended parallel sampling to the scale of millions, achieving unprecedented results in competition-level coding tasks.

However, in automated Verilog code generation, systematic studies on parallel scaling are still missing. Existing work has shown that small-scale repetition can outperform single-pass inference [27, 31]. However, the *scalability* and *effectiveness* of expanding the parallel scale to hundreds of samples remains unknown; and the underlying reasons for such improvements have yet to be explored. To this end, in this paper, we present the first work on exploring the parallel scaling in automated Verilog code generation with LLMs. Specifically, the contributions are as follows:

- We show that parallel scaling is cost-effective in both time and monetary terms, as increased concurrency substantially reduces wall-clock latency and sampling cost.
- We demonstrate that expanding the sampling size leads to consistent and significant performance improvements, surpassing even advanced reasoning-oriented LLMs such as OpenAI o1 and Gemini-2.5-pro.
- We investigate the reasons underlying these improvements and show that parallel scaling enhances both diversity and reliability of outputs.
- We analyze the influence of output randomness on parallel scaling, and find that regulating randomness preserves accuracy while further supporting convergence in large-scale sampling.

---

[1] Image generated by ChatGPT.

As shown in Fig. 1.1, the name "VerilogMonkey" is inspired by the Infinite Monkey Theorem [33], which states that given infinite time and attempts, a monkey randomly typing on a keyboard would eventually produce the complete works of Shakespeare. Similarly, LLMs can be thought of as "thinking monkeys," and with the help of parallel scaling, their potential can be further unlocked, leading to promising improvements in task performance.

## 1.2 Related Work

**Test-time Scaling**

Test-time scaling enhances task performance by allocating additional computational resources during inference. Depending on how the model's computation is expanded, test-time scaling can be divided into four categories [43]: parallel, sequential, hybrid, and internal.

*Parallel scaling* enhances performance by generating multiple outputs from LLMs concurrently and then aggregating them into a final answer. Its effectiveness depends on two metrics [6]: *coverage*, the probability of producing at least one correct output, and *precision*, the ability to identify that correct response. Empirical evidence from [6] reveals a log-linear relationship between coverage and computational resources. In tasks like competition-level programming or travel planning, automatic correctness checks are often impractical. Consequently, research has shifted toward more advanced aggregation techniques, such as majority voting [12, 24] and genetic algorithms [22].

*Sequential scaling* generates outputs iteratively, leveraging intermediate results at each stage. Techniques such as Chain-of-Thought [39] decompose complex problems into a sequence of reasoning steps. Subsequent research [9, 29] has shown that LLMs can self-correct through iterative revision. Approaches like ReAct [42] integrate external tools to provide feedback that guides the model's next action.

*Hybrid scaling* leverages the complementary strengths of parallel and sequential approaches: parallel scaling broadens an LLM's exploration of possibilities, while sequential scaling enables deeper investigation along promising paths. Notably, methods like Tree-of-Thoughts [41], Graph-of-Thought [3], and Forest-of-Thought [4] have successively introduced ever more complex hybrid scaling patterns.

*Internal scaling* empowers LLMs to allocate extra computational resources at inference time through targeted training or fine-tuning. In this paradigm, the model's internal parameters instead of external agentic mechanisms govern scaling behavior. The impressive success of reasoning models like OpenAI o1 [20] and Deepseek-R1 [16] highlights the efficacy of this approach.

**Automated Verilog Code Generation**

Existing work [5, 13] on automated Verilog code generation has predominantly focused on sequential scaling. AutoChip [37] employs reflection strategies to enhance

Verilog code generation. Additionally, it employs LLM ensembling with more powerful models when lightweight models underperform, further boosting performance. VerilogCoder [17] introduces an autonomous agent with graph-planning capabilities and integrates an AST-based waveform-tracing tool to provide feedback during code debugging. HDLCORE [32] embeds HDL expertise into Chain-of-Thought prompts, guiding LLMs through step-by-step reasoning and leveraging testbench feedback for self-verification. Prior studies have shown that performance improves as the number of repetitions increases [27, 31]. However, these efforts have been confined to small-scale repetitions, and whether scaling the sampling size to much larger magnitudes (e.g., hundreds of attempts) yields further gains remains an open question.

### 1.3 Deployment

This section explains our deployment of VerilogMonkey to support parallel scaling. VerilogMonkey focuses on generating *self-contained* [31] Verilog code, that is, implementations that do not require external module instantiations, from natural language specifications.

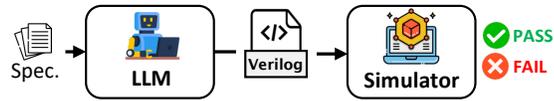

Fig. 1.2: Workflow of a single sampling iteration.

Given a specification and its associated test cases, the LLM's task is to produce functionally correct code that passes every test. Fig. 1.2 shows the workflow of a single sampling iteration: the specification is fed into the LLM, which generates the Verilog code; this code is then verified by simulation for functional correctness. VerilogMonkey repeats these iterations until either a generated implementation passes simulation or a predefined sampling limit is reached.

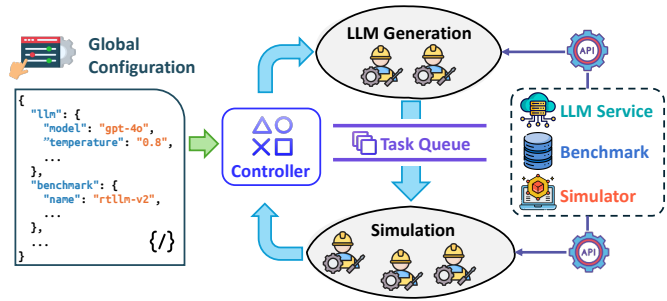

Fig. 1.3: System architecture of VerilogMonkey.

Fig. 1.3 shows the overall architecture of VerilogMonkey. The system is fully automated, and users can tailor its behavior via a global configuration file. To maximize extensibility, VerilogMonkey adopts a modular, plug-in architecture. For LLM integration, we adopt the LLM-as-a-Service (LLMaaS) paradigm, relying on providers such as OpenAI and Anthropic to perform inference. VerilogMonkey interacts with these services by sending network requests. This paradigm abstracts LLM inference as an external service, since the internal details lie beyond our scope. Tools like Ollama [30] also allow locally hosted models to be exposed as LLMaaS endpoints. We use LangChain [7] as middleware to manage these connections. On the benchmarking side, VerilogMonkey supports existing test suites [28, 31, 38] through a unified interface that eases future expansion. For simulation, it is compatible with major EDA tools, including VCS [36], ModelSim [34], and Icarus Verilog [40].

All samplings run in parallel, allowing VerilogMonkey to fully leverage concurrency and maximize throughput. The system consists of two stages: LLM generation and simulation. The LLM generation stage is I/O-bound: after dispatching inference requests, VerilogMonkey must wait for responses. To maximize utilization during this idle time, it uses batch invocation together with asynchronous processing. In contrast, the simulation stage is computation-intensive, so VerilogMonkey adopts a multi-process strategy to fully leverage all available CPU cores. These two stages are connected by a task queue, which enables pipelining and synchronization.

## 1.4 Empirical Study of Parallel Scaling

### 1.4.1 Research Questions

In this section, we first evaluate the effectiveness of parallel scaling, and analyze the reasons behind the improvements. We then identify output randomness as a key influencing factor and explore how its regulation impacts effectiveness. Accordingly, Sections 1.4.3 to 1.4.6 address the following research questions (RQs), respectively.

*RQ1.* Is parallel scaling cost-effective in terms of time and money?
*RQ2.* How effective is parallel scaling in improving performance?
*RQ3.* What are reasons for improvements?
*RQ4.* How does LLM's randomness influence effectiveness?

**Highlights**

Presented below are brief highlights of the research questions, with further details provided in the subsequent sections:

(*RQ1.*) Parallel scaling is cost-effective in both time and money. Greater concurrency sharply reduces per-job runtime, so total wall-clock time grows far more slowly than the sample size. On cost, hundreds of samples per prompt typically run $0.04–$10, and provider features like batching and prompt caching can often cut that roughly in half. (***RQ2.***) By scaling up the sampling size to 512, the LLM delivers substantial performance gains, achieving success rates on all benchmarks that outperform existing methods—including reasoning models such as OpenAI o1 and Gemini-2.5-pro. (***RQ3.***) For simple problems, LLMs produce more deterministic and confident outputs. However, errors occur due to hallucinations. Conversely, on complex or math-related tasks, their performance drops and output uncertainty grows, making small-scale repetition unreliable. In both scenarios, parallel scaling proves beneficial by enhancing the diversity and coverage of generated outputs. (***RQ4.***) Raising the model's randomness (via temperature adjustments) does not compromise single-pass accuracy and can further improve the LLM's capabilities and convergence speed in large-scale sampling.

### 1.4.2 Evaluation Setup

#### Configuration

- *Benchmark.* We evaluate parallel scaling using three of the most widely used benchmarks: VerilogEval [31], RTLLM (v2) [28] and GenBen [38]. Together, they comprise approximately 200 Verilog code generation problems, each providing a specification, test cases, and a human-written reference implementation.
- *Model.* Our evaluation includes six mainstream LLMs: Google's Gemini-2.0-flash and Gemini-1.5-flash, OpenAI's GPT-4o (20240806) and GPT-4o-mini (20240718), and Anthropic's Claude-3.7-Sonnet (20250219) and Claude-3.5-Haiku (20241022).
- *Baseline.* We compare against two categories of baselines: existing works (see Table 1.2) and reasoning models. Reasoning models include OpenAI's o3-mini and o1, Anthropic's Claude-3.7-Sonnet-Think, and Google's Gemini-2.5-pro.
- *Implementation Details.* We set the maximum sampling size to 512. For each problem, we generate samples via zero-shot prompts containing only the specification and formatting instructions to facilitate code extraction. Unless otherwise noted, all LLM inferences use default configurations (e.g., temperature, top–$p$).

#### Metrics

Our goal is to evaluate LLMs' Verilog code-generation capability in terms of success rate, i.e., the proportion of problems they can solve. We focus on two widely used metrics: Hit@$k$ [24] and Pass@$k$ [8]. Hit@$k$ is the empirical fraction of problems for which at least one of $k$ generated candidates is correct:

$$\text{Hit@}k = \frac{1}{q}\sum_{i=1}^{q}\text{hit}_i = \frac{1}{q}\sum_{i=1}^{q}\max_{1 \le j \le k}\text{Pass}(r_{i,j}), \tag{1.1}$$

where $\text{Pass}(r_{i,j}) = 1$ if the $j$-th sample for problem $Q_i$ is correct and 0 otherwise. Pass@$k$ estimates, for each problem, the probability of solving it within $k$ attempts under a Bernoulli model; when $N$ samples are collected per problem and $c_i$ of them are correct, a standard estimator is

$$\text{Pass@}k = \mathbb{E}_i \left[ 1 - \frac{\binom{N-c_i}{k}}{\binom{N}{k}} \right]. \tag{1.2}$$

The key observation is that although Pass@$k$ generally provides more robust estimates than Hit@$k$, directly computing Pass@$k$ becomes computationally impractical as $k$ grows under high concurrency. Consequently, while prior work commonly reports Pass@$k$, in our parallel-scaling setting we report Hit@$k$ instead, as it remains tractable at large $k$.

### 1.4.3 Is Parallel Scaling Cost-Effective in Terms of Time and Money?

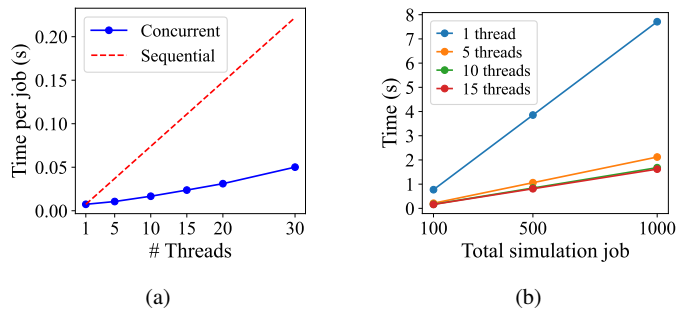

(a)                                          (b)

Fig. 1.4: Simulation performance. (a) Per-thread time to complete a job vs. concurrency (blue), compared to sequential execution (red). (b) Total time for varying numbers of jobs across thread counts.

*Time*. Fig. 1.4 shows the simulation performance under different concurrency. As concurrency increases, the average time per job drops dramatically. Therefore, the total time grows far more slowly than the sampling size. Meanwhile, LLM inference is becoming faster: for example, OpenAI's GPT-4o delivers a sixfold speedup over GPT-4 within a single year [2], and techniques like quantization and distillation have made local LLM deployments far more lightweight [45].

*Money*. Table 1.1 shows that sampling each prompt hundreds of times incurs only $0.04 to $10 in API fees, depending on the model and problem complexity. Techniques like non–real-time batch APIs and prompt-caching offered by providers can further cut these expenses by at least 50%. Even at large scale, LLM-based sampling remains far more economical than human labor: a U.S. hardware engineer earns roughly $11,000 per month [10] and may spend hours or even days to resolve a single problem.

Table 1.1: Average per-question LLM API cost (USD) for 500 samples across benchmarks.

|                   | VerilogEval [31] | RTLLM [28] | GenBen [38] |
| ----------------- | ---------------- | ---------- | ----------- |
| Gemini-1.5-flash  | 0.04             | 0.07       | 0.15        |
| Gemini-2.0-flash  | 0.06             | 0.12       | 0.34        |
| GPT-4o-mini       | 0.16             | 0.23       | 0.57        |
| GPT-4o            | 1.87             | 3.12       | 7.80        |
| Claude-3.5-Haiku  | 0.61             | 1.01       | 2.50        |
| Claude-3.5-Sonnet | 2.36             | 4.38       | 9.36        |

### 1.4.4 How Effective is Parallel Scaling?

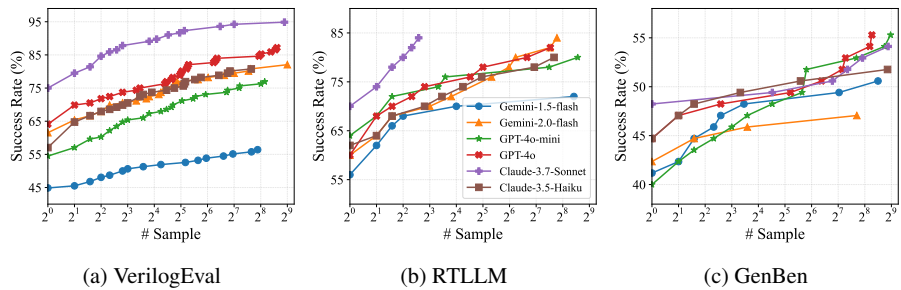

(a) VerilogEval            (b) RTLLM            (c) GenBen

Fig. 1.5: Performance improvement with increased samplings using VerilogMonkey across various benchmarks. Each point indicates a sampling where at least one problem passed the test for the first time. Success Rate (%) is measured using Hit@$k$.

Fig. 1.5 shows how success rates evolve across different benchmarks as the number of samples increases. For example, on VerilogEval, Claude-3.5-Haiku achieves a 23.72% improvement, while Gemini-2.0-flash sees a 24% jump on RTLLM. Success rates climb rapidly at first but taper off as sample size grows, following a sub-logarithmic trend ($c \propto \log \log N$): an exponential increase in samples yields logarithmic improvements.

Table 1.2 compares the success rates of various LLMs using parallel scaling against existing methods and reasoning models. VerilogMonkey outperforms all of them by simply increasing the sampling size to the hundreds, without relying on additional enhancements such as post-training or agentic mechanisms.

### 1.4.5 What Are Reasons for Improvements?

This section is organized as follows: Section 1.4.5.1 analyzes the relationship between the effectiveness of parallel scaling and two key factors influencing LLM performance, namely problem difficulty and type. Building on this analysis, we observes that this

Table 1.2: VerilogMonkey's performance across benchmarks and its comparison with baselines.

| Baseline | VerilogEval (%) | | | RTLLM v2 (%) | | | GenBen (%) *** | | |
|---|---|---|---|---|---|---|---|---|---|
| | Pass@1 | Pass@5 | Pass@10 | Pass@1 | Pass@5 | Pass@10 | | Pass@1 | Pass@5 | Pass@10 |

*Existing Work \**

| CraftRTL [26] | 68.00 | 72.40 | 74.60 | HDLCORE [32] | 60.00 | 62.00 | | GPT-3.5-Turbo | | 26.70 | - |
| OriGen [11] | 51.40 | 58.60 | 62.20 | **RTLLM v1 (%) \*\*** | | | | QWEN-v1-max | | 21.20 | - |
| AutoVCoder [13] | 48.50 | 55.90 | - | CraftRTL [26] | 49.00 | 65.80 | 74.50 | GLM-4v-plus | | 7.50 | - |
| CodeV [44] | 53.20 | 65.10 | 68.50 | OriGen [11] | | 65.50 | - | Llama3 | | 6.90 | - |
| VerilogCoder [18] | **94.20** | - | - | CodeV [44] | 36.60 | 53.30 | 61.30 | | | | |

*Reasoning Models*

| OpenAI o3-mini | 81.41 | 89.10 | 89.10 | | 72.00 | 78.00 | 78.00 | | 44.71 | 47.06 | 47.06 |
| OpenAI o1 | 74.36 | 85.90 | 85.90 | | 64.00 | 76.00 | 78.00 | | 43.53 | 47.06 | 47.06 |
| Gemini-2.5-pro | 76.92 | 90.38 | 90.38 | | 70.00 | 78.00 | **80.00** | | 45.88 | 47.06 | 47.06 |
| Claude-3.7-Sonnet-Think | 80.13 | 89.10 | 92.31 | | 68.00 | 78.00 | 78.00 | | 45.88 | 48.24 | 48.24 |

| **VerilogMonkey** | Hit@1 | Hit@512 (Improvement) | | Hit@1 | Hit@512 (Improvement) | | | Hit@1 | Hit@512 (Improvement) | |
|---|---|---|---|---|---|---|---|---|---|---|
| Gemini-1.5-flash | 44.87 | 56.41 (+11.54) | | 56.00 | 72.00 (+16.00) | | | 42.35 | 47.06 (+4.71) | |
| Gemini-2.0-flash | 61.54 | 82.05 (+20.51) | | 60.00 | **84.00 (+24.00)** | | | 41.18 | 50.59 (+9.41) | |
| GPT-4o-mini | 54.49 | 76.92 (+22.43) | | 64.00 | 80.00 (+16.00) | | | 40.00 | **55.29 (+15.29)** | |
| GPT-4o | 64.10 | **87.18 (+23.08)** | | 60.00 | **82.00 (+22.00)** | | | 44.71 | **55.29 (+10.58)** | |
| Claude-3.5-Haiku | 57.05 | 80.77 (+23.72) | | 62.00 | 80.00 (+18.00) | | | 44.71 | 51.76 (+7.05) | |
| Claude-3.7-Sonnet | 75.00 | **94.87 (+19.87)** | | 70.00 | **84.00 (+14.00)** | | | 48.24 | 54.12 (+5.88) | |

† VerilogMonkey's top-2 performances are highlighted in yellow. Best performances in baselines on each benchmark are highlighted in gray.
\* Data for the Existing Work is directly sourced from the corresponding papers. "-" indicates unavailable or not reported.

relationship further affects the ways in which the effectiveness of parallel scaling is manifested. Accordingly, Section 1.4.5.2 first uses code dispersion to summarize these different forms, and then Section 1.4.5.3 illustrates them in detail with several representative problem cases.

### 1.4.5.1 Factors Affecting Effectiveness

**Problem Difficulty**

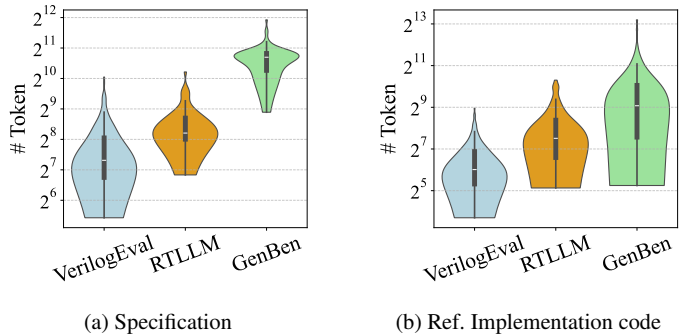

(a) Specification      (b) Ref. Implementation code

Fig. 1.6: Token length distribution for specifications and golden codes.

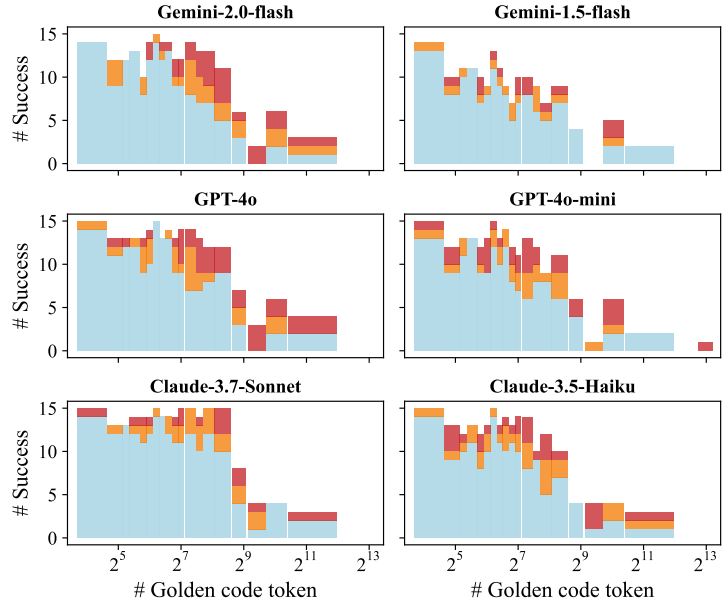

Fig. 1.7: Success rates by code length. Each bar contains 15 problems of consecutive code length. Bar width shows code length range; height shows cumulative successes. Blue represents successes at $N = 1$, orange at $N = 10$, and red at $N = 512$.

We use the length of a problem's specification or reference implementation code as an indicator for difficulty, and analyze how this correlates with success rates. Fig. 1.6 shows the token length distribution for both specifications and reference implementation codes. Specification token lengths are measured using GPT-4o's tokenizer; code token lengths are measured after parsing with pyVerilog. Across all three benchmarks, token lengths grow progressively, consistent with declining success rates reported in Table 1.2. Fig. 1.7 further examines the distribution of success rates as a function of code length. We sort problems by token count, group them into bins of 15, and display each group as a bar: the bar's width shows the token-length range, and its height indicates the number of successes. Within each bar, the blue, orange, and red segments denote successes at $N = 1$, $N = 10$ and $N = 512$, respectively. The figure reveals that model performance degrades as problem difficulty increases. For simpler problems (token count below $2^6$), LLMs already achieve high success rates under single-sampling. For moderately difficult problems (token count between $2^6$ and $2^{10}$), the $N = 1$ success rates drops, while parallel sampling yields the largest improvements. However, for the hardest problems (token count exceeding $2^{10}$), LLMs struggle to generate correct solutions, and the benefits of parallel sampling become limited. This trend suggests that parallel sampling's benefits arise from more effectively leveraging the LLMs' reasoning capabilities, rather than bypassing them. Ultimately, the model's inherent capacity sets the upper bound on achievable success.

**Problem Type**

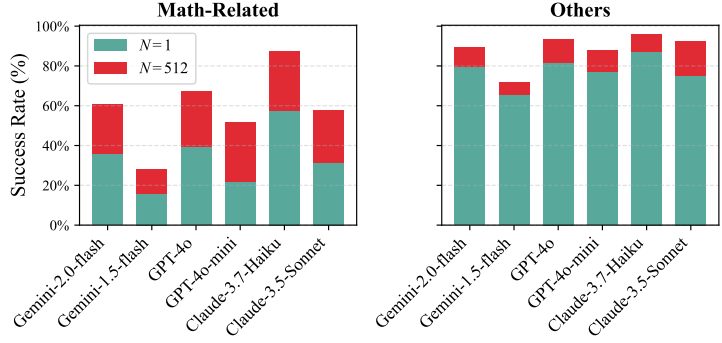

Fig. 1.8: Parallel scaling contribution for different type of problems on the VerilogEval benchmark.

The VerilogEval benchmark includes a special category of math-related tasks—finite state machines (FSMs), waveform-based logic analysis, and Karnaugh maps—comprising 40% of the entire problem set. Although these tasks are not especially complex, LLMs still exhibit uncertain performance due to their inherent challenges with math reasoning and formal derivation [1]. Fig. 1.8 shows the success rates for both categories, demonstrating that LLMs fare significantly worse on math-related problems. However, parallel scaling delivers substantial improvements for these problems. All LLMs except Gemini-1.5-flash achieve more than a 20% increase in success rates.

### 1.4.5.2 Manifestations of Effectiveness

At a high level, parallel scaling broadens the diversity and coverage of generated attempts, thereby eliminating single points of failure and increasing the likelihood of a correct solution. These benefits become more nuanced at finer granularity and are directly related to the dispersion of LLM outputs. To quantify this dispersion, we measure the pairwise similarity among all sampled codes. Specifically, for a given set of sampled codes, each code is first tokenized using pyVerilog's parser. The resulting token lists are vectorized using TF-IDF over N-grams. After that, getting $N$ code vectors $\{v_1, \ldots, v_N\}$, the cosine similarity between any two vectors $v_i$ and $v_j$ is

$$\cos \theta_{i,j} = \frac{v_i \cdot v_j}{\|v_i\|_2 \|v_j\|_2}.$$  (1.3)

The mean cosine distance is then given by

$$\text{MCD} = \frac{2}{N(N-1)} \sum_{1 \le i < j \le N} \left(1 - \cos\theta_{i,j}\right). \tag{1.4}$$

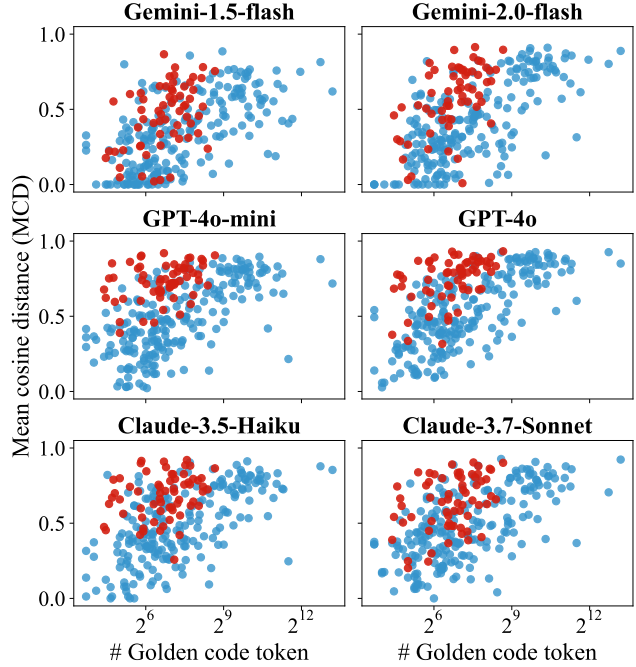

Fig. 1.9: Scatter plots of golden code lengths versus code MCDs. Math-related problems in VerilogEval are highlighted in red.

A higher MCD indicates greater dispersion among the sampled codes. Fig. 1.9 shows the scatter distribution of the reference implementation code length and the MCD of sampled codes for each problem across different models. Math-related problems in VerilogEval are highlighted in red. Overall, code length correlates positively with dispersion; and notably, math-related problems exhibit higher dispersion than other tasks of comparable length. Building on Fig. 1.9 and our analysis of Fig. 1.7, we identify two principal ways in which parallel scaling drives improvement:

- LLMs perform better on simple problems, producing outputs that are more deterministic and confident, resulting in a more concentrated code distribution. However, errors still arise from hallucinations. By repeated attempts, the probability of generating code without hallucinations increases.
- For complex and math-related problems, LLM performance decreases. Output uncertainty rises, leading to a more dispersed code distribution. As a result, small-scale repetition becomes unreliable. With parallel scaling, the LLM can attempt various implementations across many samples, thereby increasing the chances of finding a correct solution.

### 1.4.5.3 Case Study

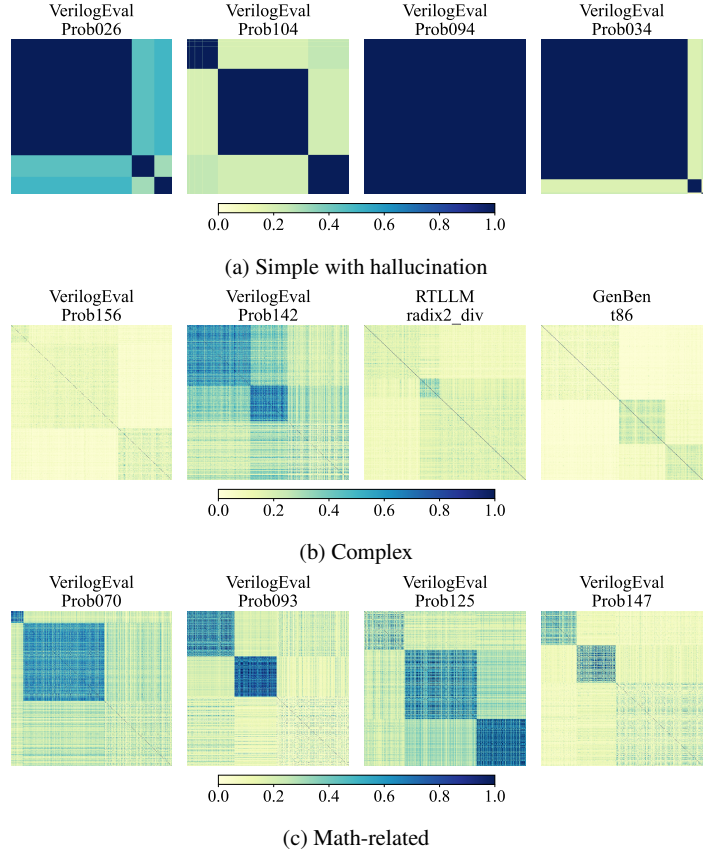

(a) Simple with hallucination

(b) Complex

(c) Math-related

Fig. 1.10: Similarity heatmap of different types of problems.

We select several representative problems across different types and collect all failed code attempts to generate the similarity heatmaps shown in Fig. 1.10. For each problem, we construct a heatmap $H$ in which the intensity of cell $H[i, j]$ indicates the textual similarity between the $i$-th and $j$-th code samples, computed as $\cos\theta_{i,j}$ in (1.3). Since $\cos\theta_{i,j}$ is symmetric, $H[i, j] = H[j, i]$. To more intuitively highlight clustering tendencies among the code samples, we apply KMeans clustering and then reorder the samples by their cluster labels. As a result, the heatmaps exhibit block-like patterns, with each block reflecting the high intra-cluster similarity. We now interpret the two manifestations of improvements.

**Hallucination**

In our context, we define an LLM "hallucinates" when it confidently generates answers that are actually incorrect [25]. Fig. 1.10a illustrates several problems where such hallucinations occur. In these cases, the model repeatedly produces only a small set of distinct outputs, even after hundreds of attempts. For example, in the VerilogEval benchmark, problem `Prob026` yields just three different code variants, and every attempt on `Prob094` produces the exact same output. Once an LLM becomes trapped in these erroneous patterns, it is extremely unlikely to escape under small-scale repetition. Scaling up to hundreds of parallel attempts can increase the odds of obtaining a correct (non-hallucinated) output, but the effectiveness of this strategy depends on how strongly the hallucination dominates. In the four problems shown, only one or two out of 512 attempts succeed, and in some cases the model never escapes the hallucination.

While most hallucinations arise from the LLM's intrinsic limitations, some are actually triggered by ambiguous or misleading specifications. For example, in the RTLLM benchmark, the task `clkgenerator` asks for a module that outputs a periodically toggling square-wave clock signal. However, because the prompt is phrased as a purely behavioral description, any code produced cannot be synthesized or simulated. The specification neither requires an RTL-level implementation nor defines the necessary input signals, as a frequency divider must receive an existing clock input. In fact, even the provided reference code is behavioral and therefore incorrect. A similar issue appears in `Prob112` of the VerilogEval benchmark. These cases highlight the critical importance of clear, precise specifications for fair and objective evaluation.

**Uncertainty**

Fig. 1.10b and 1.10c present similarity heatmaps for complex and math-related problems, respectively. In contrast to the tight clusters seen in Fig. 1.10a, these heatmaps display far greater dispersion and much weaker clustering. Such uncertainty makes small-scale repetition unreliable, since it cannot cover the full range of possible outputs. Moreover, in some cases the correct solution is not the model's most confident prediction. For example, in VerilogEval's `Prob125` and `Prob093`, the largest clusters consist of incorrect code variants. Parallel scaling, i.e., generating many attempts in parallel, does improve the odds of finding a correct answer, but this benefit is ultimately bounded by the model's intrinsic capabilities. For problems of high difficulty, the gains from parallel scaling quickly diminish (see Fig. 1.7).

### 1.4.6 How Does LLM's Randomness Influence Effectiveness?

Since parallel scaling boosts an LLM's performance by broadening its output coverage, we naturally ask: *can we further improve results by increasing output randomness*? The answer is yes. The temperature parameter [14] controls this randomness. Lower temperatures yield more deterministic, repeatable outputs, while higher temperatures encourage diversity. We demonstrate this effect on two Gemini models, which support temperatures from 0 to 2.0. Fig. 1.11 plots the distribution of code mean cosine distances (MCDs) at three temperature settings, and Fig. 1.12 shows how success rates evolve with sample size under each temperature. At low temperatures, outputs vary little and success rates quickly plateau. For instance, Gemini-2.0-flash shows no improvement beyond five samples when $T = 0$. As we raise the temperature, LLMs achieve higher final success rates and converge faster. However, too much randomness can be detrimental: for GPT-4o and GPT-4o-mini, temperatures above 1.8 (on a 0–2.0 scale) rarely produce coherent text. Importantly, increasing randomness does not harm accuracy when sampling is small. With a single sample ($N = 1$), success-rate differences across temperatures remain within 5%. We attribute this variance primarily to the inherent noise of individual runs.

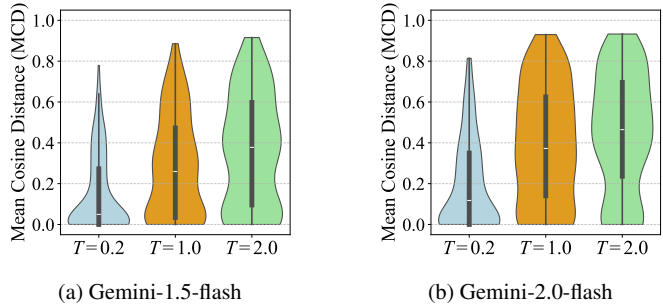

Fig. 1.11: Code MCD distributions at different temperature settings.

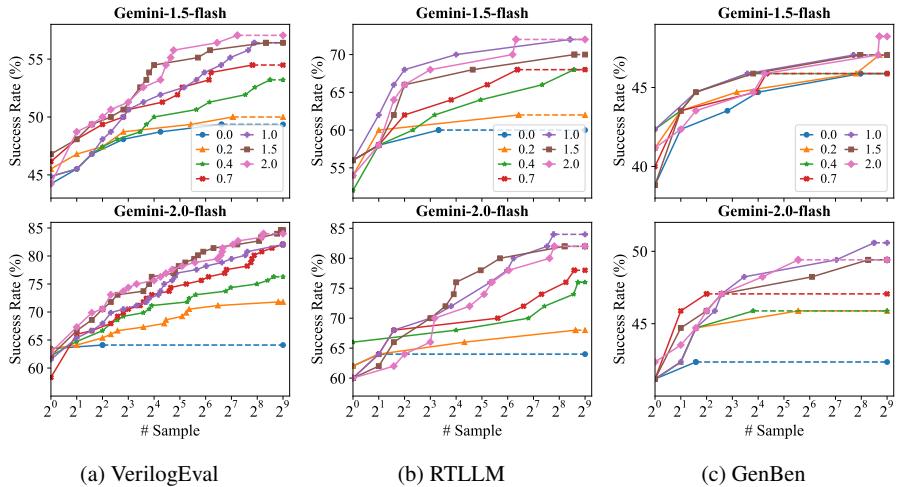

Fig. 1.12: The impact of temperature on success rate. Success Rate (%) is measured using Hit@$k$.

## 1.5 Conclusion

We present VerilogMonkey, an empirical study of parallel scaling for automated Verilog generation. Scaling to hundreds of parallel samples—without enhancements like post-training or agentic methods—is cost-effective in time and money and surpasses prior LLM-based results across benchmarks and mainstream models. Our analysis explains these gains, showing that parallel scaling improves diversity and reliability, and that controlling output randomness shapes its effectiveness.